# Modeling and Nonlinear Control of Gantry Crane Using Feedback Linearization Method

Ehsan Omidi
ehsaneomidi@gmail.com

*Abstract*—Due to the requirements of high positioning accuracy, small swing angle, short transportation time, and high safety, both motion and stabilization control for an gantry crane system becomes an interesting issue in the field of control technology development. In this paper, dynamic model of gantry crane is extracted using Lagrange method. This model has been linearized and weaknesses of state feedback control of this model is reviewed. To solve these problems, state feedback gain matrix is calculated using LQR method and results are fully investigated. Finally, a full nonlinear solution of problem using feedback linearization method is implemented and results are compared with previous works.

## I. INTRODUCTION

Gantry crane systems are widely used in harbors and factories for loading and unloading of goods. The crane systems are desired to be able to move to the required positions as fast and as accurate as possible while placing the payload at the appropriate position [1]. In the context of further automation of manufacturing processes, automated transportation of heavy weights using cranes becomes more and more important. Applying the skills of robots to crane automation, a wide market of new applications could be developed [2]. Till now cranes are one of the most important systems for material handling of heavy goods. Although automatic cranes are comparatively rare in the industrial practice. Because of the high potential of rationalization, in the past several attempts have been made. But, several reasons prevented the success of such systems. Till now, one of these reasons is the relation between investment costs and achievable cost savings. But, due to decreasing investment costs because of lower prices for hard- and software as well as for actuators and sensors the profitability of such systems is within reach [3]. The need for faster cargo handling requires control of the crane motion so that its dynamic performance is optimized [4].

The first idea in crane automation was calculating time optimal control functions minimizing the traveling time of the crane considering the boundary conditions of no load swaying at the target point [4]. Assuming idealized constraints the result was a feedforward bang–bang-controller for the force on the crane trolley. This approach was further developed by Auernig and Troger [5], Manson [6], Sakawa and Shindo [7]. For industrial use, however, several restrictions are necessary [8]. The maximum acceleration must be limited to avoid exciting unmodeled eigen frequencies. Thus, the control needs a feedback loop to guarantee damping of load oscillations [9]. In order to increase robustness in industrial applications, one has to give up the requirement of time optimality and calculate the control function just under the boundary conditions of angle and make sure that the angular velocity is zero at the target point [10]. Also, feedforward control alone is not able to damp oscillations due to disturbances, or initial conditions of the rope angle. Therefore, these approaches are only suitable in case of milder requirements on positioning and tracking accuracy. These systems are useful to increase operational safety and to save costs by downgraded employee training [3]. Sliding mode control is also one of the most frequently used approaches in terms of robustness. It has been used in controlling the spatial cable robot [11], and also in high precision positioning tasks [12, 13], where uncertainty plays an important role.

Higher positioning and tracking accuracy is only achieved by feedback control. The main problems for control design are determined by the nonlinear dynamic system behavior, the time variance of the system, as well as the coupling of the dynamics between the different moving directions of the crane. In this study, the dynamical behavior of a gantry crane is investigated through the extraction of the model by means of the Lagrange method. After a linear estimation of the model, a state feedback control method is used and the full nonlinear model is tested with that controller. In order to have a better state feedback controller and also acceptable response for the nonlinear model, feedback gain matrix is calculated by LQR method to maximize the validation of linear model for nonlinear system. After designing an observer for immeasurable states of system in real world because of some issues, a transformation on states of system is executed using feedback linearization method and the result is a system which can be controlled by and linear method,

E. Omidi. is with the Department of Mechanical Engineering of Iran University of Science and Technology, (phone: +989126422606; e-mail: ehsanomidi@mecheng.iust.ac.ir).

whereas no nonlinear aspect of the problem is removed.

## II. DYNAMICAL MODEL

### A. Calculating the Equations of Motion

The primary use of gantry cranes is to move heavy objects. A gantry crane has a trolley moving on a track, and the object to be moved is hung by a rod. Fig. 1 shows the crane coordinate and parameters, where x is the trolley position, θ is the rod angle, l is the rod length, m is the trolley mass, M is the payload mass, and F(t) is the force applied to the trolley.

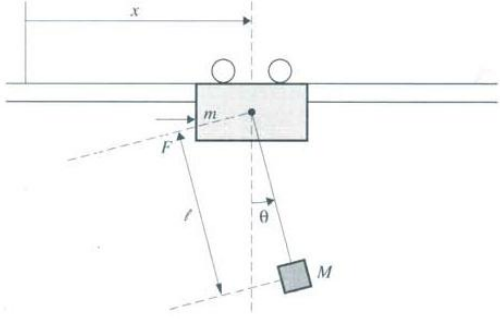

Fig. 1. Gantry crane system

According to the steps of Lagrange method, the calculation of kinetic and potential energies are:

$$T = \frac{1}{2}m\dot{x}^2 + \frac{1}{2}M(\dot{x}^2 + \dot{q}^2 l^2 + 2\dot{x}\dot{q}l\cos q) \quad (1)$$

$$V = Mgl(1-\cos q) \quad (2)$$

$$L = T - V \quad (3)$$

$$\frac{d}{dt}(\frac{\partial L}{\partial \dot{q}_i}) - \frac{\partial L}{\partial q_i} = Q_i \quad (4)$$

By means of the Lagrangian equation (4), and defining the generalized coordinate $q_i$, and generalized force $Q_i$, the motion equations of the overhead crane system can be obtained with respect to the generalized coordinates x and θ:

$$(M+m)\ddot{x} + (Ml\cos q)\ddot{q} - Ml\dot{q}^2 \sin q = F \quad (5)$$

$$(\cos q)\ddot{x} + (l)\ddot{q} + g\sin q = 0 \quad (6)$$

In order to prepare the equations for state form, the accelerations are calculated as:

$$\ddot{x} = \frac{F + Ml\dot{q}^2 \sin q + Mg\sin q \cos q}{M + m - M\cos^2 q} \quad (7)$$

$$\ddot{q} = \frac{(F + Ml\dot{q}^2 \sin q)\cos q - (m+M)g\sin q}{l(M+m-M\cos^2 q)} \quad (8)$$

By defining state parameters such that $x_1 = x$, $x_2 = q$, $x_3 = \dot{x}$ and $x_4 = \dot{q}$, the state equations are as follows:

$$\dot{x}_1 = x_3 \quad (9)$$

$$\dot{x}_2 = x_4 \quad (10)$$

$$\dot{x}_3 = \frac{F + Mlx_4^2 \sin x_2 + Mg \sin x_2 \cos x_2}{M + m - M\cos^2 x_2} \quad (11)$$

$$\dot{x}_4 = \frac{(F + Mlx_4^2 \sin x_2)\cos x_2 - (m+M)g\sin x_2}{l(M+m-M\cos^2 x_2)} \quad (12)$$

In order to have a physical sense of dynamical behavior of the system, simulation of extracted equations for the input force of 1 kN is shown in Fig. 2. The numerical values of parameters are 500 kg for trolley, 2000 kg for pendulum and 10 m for the length of rod.

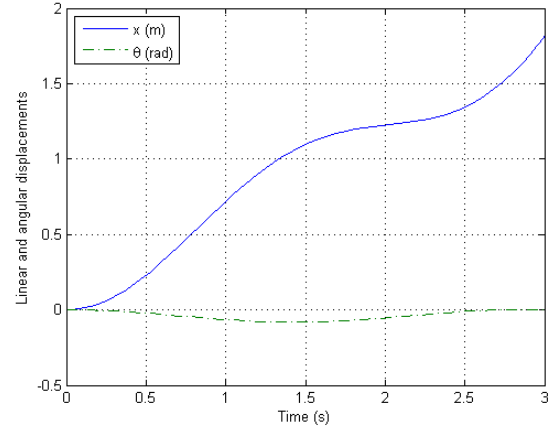

Fig. 2. Angular and linear displacements for input force of 1kN

As it is obvious in the profile of linear displacement, because of the heavy mass of the load, the changes in angular displacement is insignificant but it has a sensible effect on the trolley's move.

### B. Linearization of Equations

Assuming pendulum deviation (swing angle) is not too large, the linearized model will be a good estimation for the real gantry crane model. Either can the differential equations of motion be linearized or the mathematical model in state space. In order to have a simple perspective of the problem and the state feedback control, equations (7) and (8) are linearized according to the methods of control resources that the [14]. The resulting equations are:

$$(M+m)\ddot{x} + (Ml)\ddot{q} = F \quad (13)$$

$$\ddot{x} + l\ddot{q} + gq = 0 \quad (14)$$

Transfer function of this system is calculated regarding to the entrance force F(t) and the numerical values of parameters are as mentioned:

$$\frac{X(s)}{F(s)} = \frac{0.002}{s^4 + 4.9s^2} \quad (15)$$

$$\frac{q(s)}{F(s)} = \frac{-0.0002s^2}{s^4 + 4.9s^2} \quad (16)$$

In the subject of controllability of the linearized system, the controllability matrix calculated for system according to relationships from [14] is full rank and controllability of the system is ensured.

## III. LINEAR CONTROL APPROACHES

### A. State Feedback Control

The state feedback control means that there is a control $u = -k^T x$, proportional with the state vector x. Therefore, the full-state feedback control is applied to the linear system. In order to have the desired response of the linearized system and to move its poles to $p_1$, the feedback state gain of the system is calculated and the simulation result is demonstrated in Fig. 3. The assumption made is that the state vector of the gantry crane mathematical model can be completely measured.

$$p_1 = [-1, -2, -2 \pm 2i] \tag{17}$$

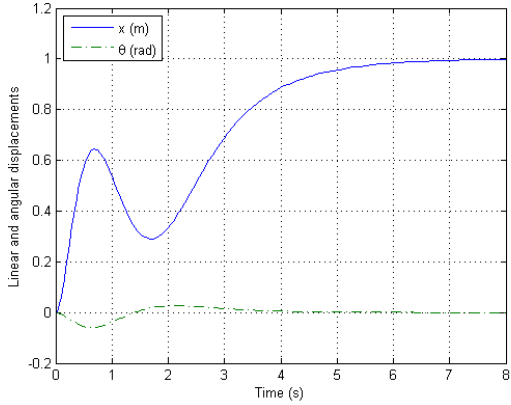

Fig. 3. Angular and linear displacements for desired final displacement of 1 m, by means of state feedback control

If there is a faster response required, using $p_2$ and by simulation, the final displacement of 5 m is achieved in a shorter time (Fig. 4) but an undesirable overshoot happens.

$$p_2 = [-4, -8, -8 \pm 8i] \tag{18}$$

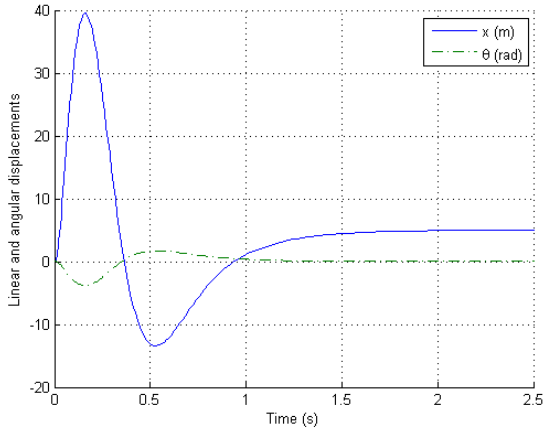

Fig. 4. Angular and linear displacements for desired final displacement of 5 m, by means of state feedback control and remote poles

Using the calculated control input according to the state feedback for nonlinear model, for displacements that do not trespass the linear bound for angle of the pendulum, a reasonable response is procured. But for larger values of final displacements of trolley, since there is a bigger input force required to keep up with the defined time limitation, the final response is completely undesirable and expresses the limitation of linearization. To solve this problem, gain matrix is calculated through a more reliable method of LQR in the following section.

### B. Calculation of Feedback Gain Matrix Using LQR Method

The main objective of this work is to design robust, fast, and practical controllers for gantry cranes to transfer the load from point to point in a short time as fast as possible and, at the same time keep the load swing small during the transfer process and completely eliminate it at the load destination. therefore by keeping swing angle as small as possible, the validation of linearized model increases to maximum point. Linear Quadratic Regulator (LQR) is used to find the optimal feedback gain matrix, whereas the intended performance is achieved. Examples of novel implementations of this approach can be found in [17, 18]. First step is to define Q and R matrixes and calculating gain matrix in a way that J is minimized. J is defined as:

$$J = \int_0^\infty \left[ x^T(t) Q x(t) + u(t)^T R u(t) \right] dt$$

(19)

Where Q and R are symmetric positive semi-definite and positive definite matrixes. For our desired performance for problem, we define the arrays of matrix Q as $Q_{11} = 1/10^2$, $Q_{22} = 1/0.2^2$ and for matrix R we have $R_{11} = 1/1500^2$. By setting these values, we prohibit x from increasing more than 10 m and swing angle from 0.2 rad. This helps the robustness of linear model. By running simulations using Matlab software and solving the problem, the resulted graph for final cargo displacement to distance of 1 m is shown in Fig. 5. The feedback gain matrix k, is also as follows:

$$k = [1, -6.52, 1.35, 9.49] \times 10^4 \tag{20}$$

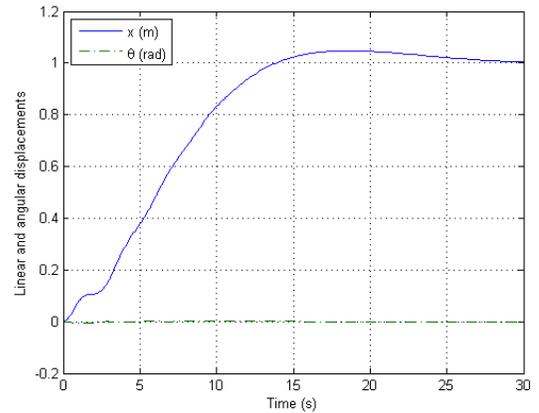

Fig. 4. Angular and linear displacements for desired final displacement of 1 m, by means of state feedback control and LQR

As it is discernable, lower angle twists and smother movement of trolley is one of achievements of this method. For investigating the lucrative performance of LQR more, we are going to use this control law for the nonlinear system, in the range, which the previous state feedback failed for larger displacements of trolley. The result for final displacement of 10 m is demonstrated in Fig. 5.

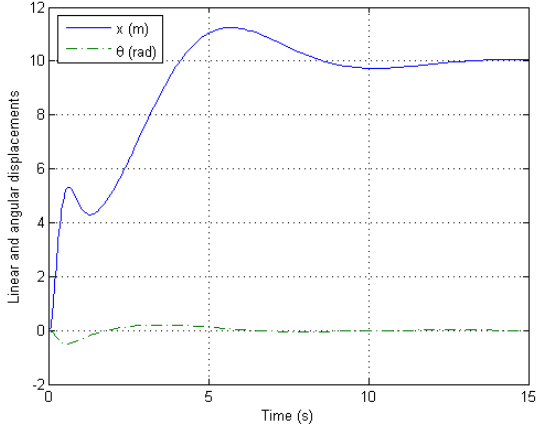

Fig. 5. Angular and linear displacements for desired final displacement of 10 m, by means of state feedback control and LQR for full nonlinear sys.

A reliable behavior of system for full model is presented and result is satisfactory enough. For next part, designing an observer is addressed since all states of system might not be measurable during the process.

*C. State Observer Design*

In this stage, we suppose that the only measurable state of system is its linear displacement. In this procedure, a third order observer is issued and performed. By means of the method related in [14], and by assuming a reasonable speed of convergence, the results of designing and simulating the observer, and its estimation for speed of trolley and swing angle are respectively demonstrated in Fig. 6 and Fig.7. As it is discernable, convergence has happened in less than 1 second after the beginning of movement, which is a good performance.

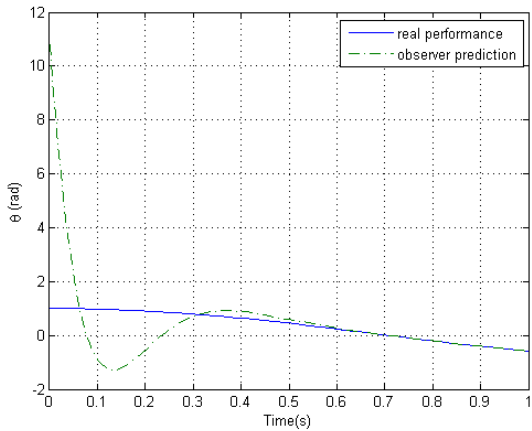

Fig. 6. Observer estimation of angular displacement

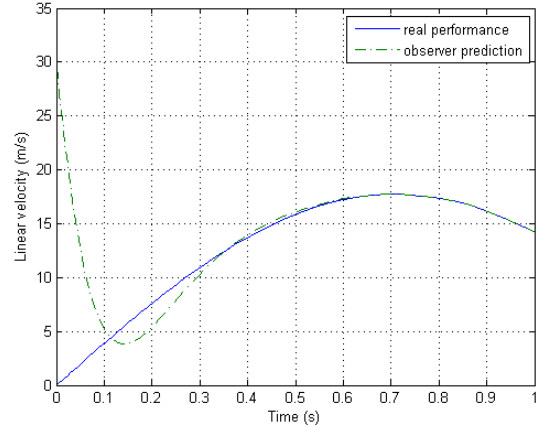

Fig. 7. Observer estimation of linear velocity of trolley

IV. NONLINEAR CONTROL SYSTEM DESIGN

In this section, a nonlinear controller design scheme based on feedback linearization method will be presented for the underactuated gantry crane system. Feedback linearization is one of the most practical control approaches for such systems, since you can transform the states of a system and form a system which could be controlled by means of a linear controller. In this method, no nonlinear behavior of system is removed and the performance has an acceptable closeness to experimental results of such gantry cranes. In Fig. 8, a schematic view of what happens through this approach is illustrated.

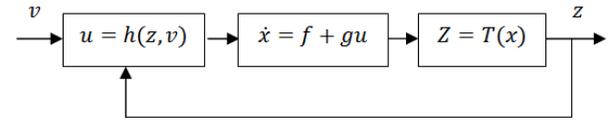

Fig. 8. The block box diagram of FBL method

As it is discernable, a transformation Z, on the states of system is performed and by implementation of resulted feedback of this transformation, a new input to nonlinear system is designed, using a linear control input of v for states named Z.

In order to perform this task for gantry crane system some complexity is added to model and they are defined and noted on Table 1 [15]. Using these additives, the resulted model through the previous Lagrange method is:

$$\dot{x}_1 = x_3 \qquad (21)$$

$$\dot{x}_2 = x_4 \qquad (22)$$

$$\dot{x}_3 = \frac{-0.5Ml.g \sin 2x_2 - Mgc_2 x_2 \cos x_2}{M + m - M \cos^2 x_2} \\ - \frac{Mlx_4^2 \sin x_2 + c_1 x_2 + f_s - F}{M + m - M \cos^2 x_2} \qquad (23)$$

$$\dot{x}_4 = \cos x_2 h(\bar{x}) - g \sin x_2 - \frac{c_2}{ml} x_4 \qquad (24)$$

for equation (24), it is notable that $h(\bar{x}) = x_3$. The following format is assumed for state space equations and f and g matrixes are calculated so:

$$\dot{\bar{x}} = f(\bar{x}) + g(\bar{x})u \quad (25)$$

In order to perform the feedback linearization task, it is necessary to check whether our system is feedback linearizable or not. According to [16], by constructing matrix $M_c$ and checking the independency of its vectors, we are assured of its conceivability.

$$M_C = \begin{bmatrix} g & ad_f g & ad_f^2 g & ad_f^3 g \end{bmatrix} \quad (26)$$

Where $ad_f g$ is the lie bracket of $f$ and $g$. after constructing the mentioned matrix and checking its rank to be full, next step is to calculate the state transformation for the desired system. The steps of finding this transformation are [16]:

$$z_1 = T_1(\bar{x}) \quad (27)$$

$$z_i = T_i(\bar{x}) = L_f T_{i-1} \quad (28)$$

And $T_1$ is calculated through following equation:

$$T_1 = (M_C^T)^{-1} \times \begin{bmatrix} 0 & 0 & 0 & c \end{bmatrix}^T \quad (29)$$

Where c has a desirable constant value. Finally the control input which is inserted to nonlinear system is:

$$u = a(\bar{x}) + b(\bar{x})v \quad (30)$$

Where α and β are:

$$a(\bar{x}) = -b(\bar{x}) L_f T_4 \quad (31)$$

$$b(\bar{x}) = \frac{1}{L_g T_4} \quad (32)$$

Following the steps as mentioned above, and finding transformation and related input of system the final linearization in procured. Using Matlab software for calculating different sections of this procedure and by using a linear PID controller with optimized coefficients, the resulted output for final displacement of 1 m for described model is achieved and demonstrated in Fig. 7. The related swing angle is also depicted in Fig. 8.

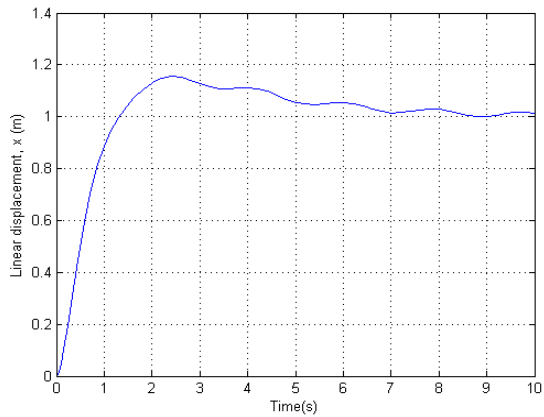

Fig. 7. Trolley's displacements for desired final value of 1 m, by means feedback linearization method and a linear PID controller.

TABLE I
UNITS FOR GANTRY CRANE PROPERTIES

| Symbol | Description | Values |
|---|---|---|
| m | Mass of trolley | 1.6 kg |
| M | Mass of swing load | 0.4 kg |
| L | Length of rod | 0.4 m |
| g | acceleration | 9.8 m/s$^2$ |
| $f_s$ | Friction force | 0.4 N |
| $c_1$ | Friction coefficient of the linear motor | 0.4 |
| $c_2$ | Friction coefficient of the rod | 0.5 |

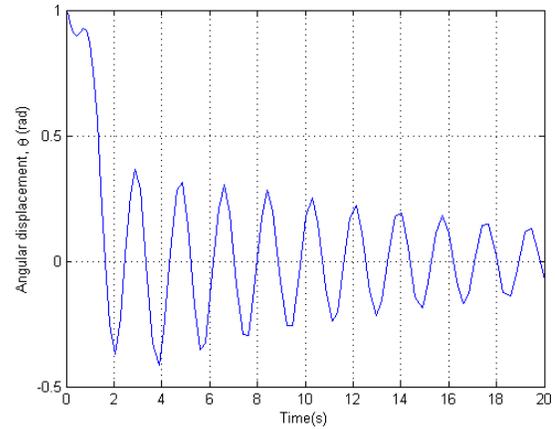

Fig. 8. Swing angle changes for desired final linear displacement of 1 m, by means feedback linearization method and a linear PID controller for initial condition of 1 rad.

## V. CONCLUSION

As it was demonstrated in this study, after extraction of a dynamical model for gantry crane, in order to get an extensive sight of the control approach, linearization of model was issued and after using a state feedback control for moving the trolley to a desirable location, we observed the incompetency of such method for larger values of x. To come up with this problem, we proposed the LQR method and finding and optimizing the gain values for state feedback controller was executed by means of this regulator. Application of controller for nonlinear model was satisfactory even for larger values that previous controller failed. In order to have a controller capable of any desired action considering the nonlinearities of a more elaborate model, feedback linearization method was proposed and performed and by means of linear PID controller, the full nonlinear model was controlled for any desirable goal values.